\documentclass{aa}
\usepackage{graphicx}
\usepackage{natbib}
\usepackage{color}
\usepackage{float}
\usepackage{txfonts}
\usepackage{pdflscape}

\begin{document}

\title{CzeV1731: The unique doubly eclipsing quadruple system }

\author{Zasche,~P.~\inst{1},
        Henzl,~Z.~\inst{2},
        Lehmann,~H.~\inst{3},
        Pepper,~J.~\inst{4},
        Powell,~B.P.~\inst{5},
        Kostov,~V.B.~\inst{5,6},
        Barclay,~T.~\inst{5},
        Wolf,~M.~\inst{1},
        Ku\v{c}\'akov\'a,~H.~\inst{1,2,7,8},
        Uhla\v{r},~R.~\inst{9},
        Ma\v{s}ek,~M.~\inst{2,10},
        Palafouta,~S.~\inst{11},
        Gazeas,~K.~\inst{11},
        Stassun,~K.G.~\inst{12},
        Gaudi,~B.S.~\inst{13},
        Rodriguez,~J.E.~\inst{14},
        Stevens,~D.J.~\inst{15,16}
        }

\offprints{Petr Zasche, \email{zasche@sirrah.troja.mff.cuni.cz}}

 \institute{
  $^{1}$ Astronomical Institute, Charles University, Faculty of Mathematics and Physics, V~Hole\v{s}ovi\v{c}k\'ach 2, CZ-180~00, Praha 8, Czech Republic\\
  $^{2}$ Variable Star and Exoplanet Section of the Czech Astronomical Society, Vset\'{\i}nsk\'a 941/78, CZ-757 01 Vala\v{s}sk\'e Mezi\v{r}\'{\i}\v{c}\'{\i}, Czech Republic \\
  $^{3}$ Th\"uringer Landessternwarte Tautenburg, Sternwarte 5, 07778, Tautenburg, Germany\\
  $^{4}$ Department of Physics, Lehigh University, 16 Memorial Drive East, Bethlehem, PA 18015, USA\\
  $^{5}$ NASA Goddard Space Flight Center, Greenbelt, MD 20771, USA\\
  $^{6}$ SETI Institute, 189 Bernardo Ave, Suite 200, Mountain View, CA 94043, USA\\
  $^{7}$ Astronomical Institute, Academy of Sciences, Fri\v{c}ova 298, CZ-251 65, Ond\v{r}ejov, Czech Republic\\
  $^{8}$ Research Centre for Theoretical Physics and Astrophysics, Institute of Physics, Silesian University in Opava, Bezru\v{c}ovo n\'am. 13, CZ-746 01, Opava, Czech Republic\\
  $^{9}$ Private Observatory, Poho\v{r}\'{\i} 71, CZ-254 01 J\'{\i}lov\'e u Prahy, Czech Republic \\
  $^{10}$ FZU - Institute of Physics of the Czech Academy of Sciences, Na Slovance 1999/2, CZ-182~21, Praha, Czech Republic\\
  $^{11}$ National and Kapodistrian University of Athens, Department of Physics, Section of Astrophysics, Astronomy and Mechanics, GR 15784 Zografos, Athens, Greece\\
 $^{12}$ Vanderbilt University, Department of Physics \& Astronomy, 6301 Stevenson Center Ln., Nashville, TN 37235, USA\\
 $^{13}$ Department of Astronomy, The Ohio State University, 140 West 18th Avenue, Columbus, OH 43210, USA\\
 $^{14}$ Center for Astrophysics \textbar \ Harvard \& Smithsonian, 60 Garden St, Cambridge, MA 02138, USA\\
 $^{15}$ Department of Astronomy \& Astrophysics, The Pennsylvania State University, 525 Davey Lab, University Park, PA 16802, USA\\
 $^{16}$ Center for Exoplanets and Habitable Worlds, The Pennsylvania State University, 525 Davey Lab, University Park, PA 16802, USA
 }

\titlerunning{The unique doubly rclipsing quadruple system CzeV1731}
\authorrunning{Zasche et al.}

  \date{Received \today; accepted ???}

\abstract{
  We report the discovery of the relatively bright ($V=10.5$~mag), doubly eclipsing 2+2 quadruple system
  CzeV1731. This is the third known system of its kind, in which the masses are determined for all four
  stars and both the inner and outer orbits are characterized. The inner eclipsing binaries are well-detached systems moving on circular orbits: pair A with period $P_A=4.10843$~d and pair B
  with $P_B=4.67552$~d. The inner binaries contain very similar components ($q\,\approx\,1.0$), making the
  whole system a so-called double twin. The stars in pair B have slightly larger luminosities and masses
  and pair A shows deeper eclipses. All four components are main-sequence stars of F/G spectral type.
  The mutual orbit of the two pairs around the system barycenter has a period of about 34 yr and an
  eccentricity of about 0.38. However, further observations are needed to reveal the overall architecture
  of the whole system, including the mutual inclinations of all orbits. This is a promising target for
  interferometry to detect the double at about 59~mas and $\Delta M_{bol} < 1$~mag.
}


\maketitle

\section{Introduction} \label{intro}

It has been 12 years since \cite{2008MNRAS.389.1630L} discovered the first doubly eclipsing system,
V994~Her. Until then, there were no stars with two distinctive photometric periods in the data known,
apart from systems like BV+BW~Dra \citep{1982AaAS...48...85G} in which the two pairs can be visually
resolved with larger telescopes. Since the discovery of V994~Her, there have been many discoveries of
star systems showing two sets of eclipses in their light curves, but only a few have been studied in
detail. A comprehensive study of this type of system was introduced by our group in
\citet{2019A&A...630A.128Z}. The latter work included a list of all currently known doubly eclipsing
systems, introduced possible resonances in these 2+2 quadruples, and also discussed the motivation for
studying these kinds of objects.

However, only a few of those candidates have been shown to orbit around each other, constituting
real 2+2 systems (about 30 such stars out of 146 systems currently known). There are even fewer
cases of systems for which the 2+2 quadruple structure is demonstrated and for which the
component stellar masses are known based on dynamical analyses from spectra. There are currently
only two such stellar systems, in which both the inner and outer orbits are known, V994~Her and
V482~Per \citep{2017ApJ...846..115T}.  We add a third example in this paper.

\section{Discovery and data}
\label{sec:data}

The system under investigation in this work is TYC 3929-724-1 (= 2MASS~J19245582+5704084 =
TIC~284482112), located at RA = 19$^h$24$^m$55.82$^s$, Dec = +57$^\circ$ 04$'$ 08.39$''$, with
$V_{max}=10.5$~mag. Its particular type of variability was discovered by amateur astronomer and
co-author Zbyn\v{e}k Henzl in spring 2019 during observations of another close target in the same
field. The system is also known as CzeV1731, which comes from the Czech Variable Stars Catalogue
\citep{2017OEJV..185....1S}, and we use the name CzeV1731 to refer to the system through the rest of
this paper. This is a typical situation, in which the discovery itself was originally done by an
amateur astronomer and is later confirmed by a professional astronomer. This is a noteworthy example of
a scientific contribution of amateur astronomers and also represents the value of exploring archival
data and the synthesis of multiple data sets.  At this point we would like to emphasize that the Czech
Variable Stars catalog\footnote{See http://var2.astro.cz/czev.php?lang=en} contains more than 2700
records of new variable stars, mostly discovered by amateur astronomers.

To rule out the possibility that the system is a visual double (two unbound eclipsing binaries at a
close projected separation), we checked that the system is not known to be visual binary in the WDS
catalog \citep{2001AJ....122.3466M}. We verified the doubly eclipsing nature of the system by compiling
existing photometric data.  The star was observed by the ground-based, wide-field photometric surveys
like "All sky automated survey for supernovae" (ASAS-SN,
\citealt{2014ApJ...788...48S,2017PASP..129j4502K}), "Super Wide Angle Search for Planets" (SWASP,
\citealt{2006PASP..118.1407P}), "Northern Sky Variability Survey" (NSVS,
\citealt{2004AJ....127.2436W}), and "Kilodegree Extremely Little Telescope" (KELT,
\citealt{2007PASP..119..923P,2018AJ....155...39O}). These data showed two distinct eclipse periods of
about 4.1085~d, and 4.6755~d that show eclipse timing variations (ETVs). We also obtained follow-up
light curves from University of Athens Observatory (UOAO) (unfiltered; see
\citealt{2016RMxAC..48...22G}), Ond\v{r}ejov Observatory ($BVRI$), the FRAM telescope (CTA-N FRAM, La
Palma, $B$ filter), and also the private observatory of co-author R.U. (unfiltered). Furthermore, Z.H.
obtained dedicated $VRI$ Cousins photometry at his private observatory in Velt\v{e}\v{z}e u Loun, Czech
Republic, using a 30 cm reflector 305/1200 equipped with Moravian Instruments MII G2-402 CCD camera and
processed in a standard way using dark frames and flat fields for reduction.  Some of this data was
used in the analysis of the system.  We provide a list of all photometric data sources used for our
analysis in Table \ref{sourcesPhotometry}, which together span over 20 years of observations.

\begin{table}[b]
  \caption{Individual photometric data sets used for our analysis.}  \label{sourcesPhotometry}
  \scalebox{0.99}{
  \begin{tabular}{c c c c}\\[-6mm]
\hline \hline
 Source        & Range (HJD-2400000) & Filter      & $\#$ of obs.    \\
 \hline
 NSVS           & 51274 -- 51627     &  $C^\dag$   & 272  \\  
 SWASP          & 53901 -- 54688     &  $S^*$      & 9565 \\
 KELT           & 55978 -- 56840     &  $\sim R^*$ & 1494 \\
 ASAS-SN        & 56080 -- 59000     &  $V$        & 1836 \\
 TESS           & 58683 -- 58954     &  $T^*$      & 7115 \\
Velt\v{e}\v{z}e & 58892 -- 58981     &  $VRI$      & 26123\\
 \hline
\end{tabular}}
 {\small Note: $^*$ The SWASP, KELT, and TESS surveys use nonstandard filters, $^\dag$ C stands for clear (unfiltered) photometry.}
\end{table}

\begin{figure}
  \centering
  \includegraphics[width=0.49\textwidth]{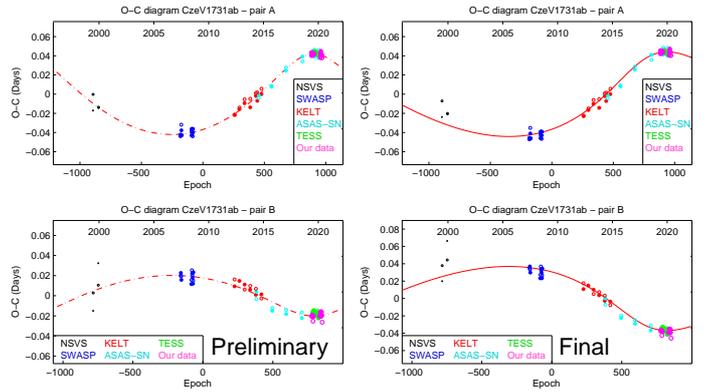}
  \caption{Analysis of period variations of both eclipsing pairs, with individual data sets labeled.
 The preliminary analysis is on the left-hand side and the final analysis is on the right-hand side of the figure. The filled dots stand
  for primary eclipses, while the open circles for the secondary eclipses. The larger the symbol, the higher
  the accuracy. See the text for details about the individual fits.}
  \label{FigOC}
\end{figure}

We conducted a preliminary analysis of the SWASP, KELT, ASAS-SN, and NSVS photometry assuming a light
ratio of the binary pairs of 50\%/50\%, using a so-called Automatic fitting procedure (AFP method,
\citealt{2014A&A...572A..71Z}). These initial results show that the observed versus predicted times for
the two photometric signals are anticorrelated on a timescale of about 26 yr; the ratio of the two ETV
amplitudes is about 2.1. This indicates a gravitationally bound, dynamically interacting pair of
binaries. See Fig. \ref{FigOC} for this preliminary solution.

\begin{figure}
  \includegraphics[width=0.45\textwidth]{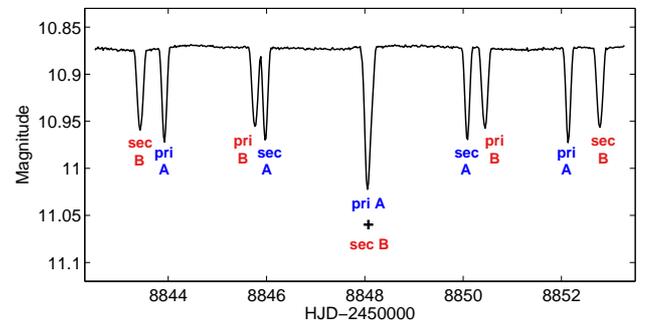}
  \caption{Segment of the TESS photometry of CzeV1731 (30 min cadence data). Eclipses of both pairs are clearly visible, in some cases overlapping.}
  \label{FigTESSdata}
\end{figure}

In addition to the ground-based photometry, the TESS mission \citep{2015JATIS...1a4003R} also observed
CzeV1731 in Sectors 14,15,16,17,20, and 23 of that mission (TESS Cycle 2). Full frame images (the FFI
data) were secured every 30 minutes, which is sufficient for our purpose. For our analysis we used the
reduced point spread function flux values. For illustration a segment of 30-minute cadence TESS light
curve from Sector 20 is plotted in Figure \ref{FigTESSdata}. These TESS data are used in the BJD time
frame, while the other ground-based photometry was transformed from Heliocentric Julian Date (HJD) to
Barycentric Julian Date (BJD).

\begin{figure}
  \centering
  \includegraphics[width=0.45\textwidth]{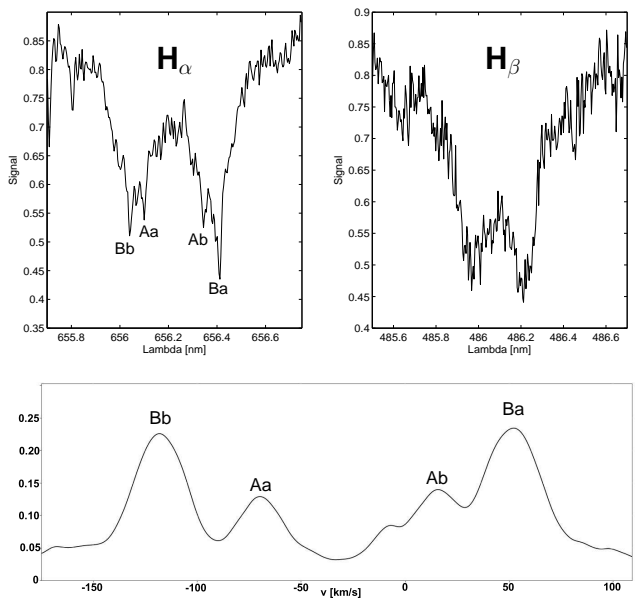}
  \caption{Two portions of the highest-quality spectrum from the Tautenburg observations, showing the
    $H_\alpha$ (upper left panel) and $H_\beta$ (upper right panel) spectral features.  When this spectrum
    was obtained on 15 September 2019, the phases of both binaries were $\Phi_A=0.088$ and $\Phi_B=0.706$.
    The lower plot shows the cross-correlation function of this spectrum. All four components are
    distinguishable in $H_\alpha$, but are hardly detectable in $H_\beta$. }
  \label{FigSpectra2}
\end{figure}

We also obtained spectroscopic observations of CzeV1731 for dynamical analysis via radial velocity
(RV) measurements. Twelve spectra were taken at Tautenburg Observatory using its two-meter telescope
equipped with an echelle spectrograph during September 2019 with an exposure time of 2400~s and a
spectral resolving power of 58000, resulting in a typical S/N ratio of 80. An example of the two
lines $H_\alpha$, and $H_\beta$ obtained out of conjunction are shown in Figure \ref{FigSpectra2},
demonstrating the visible presence of two separate pairs of blended spectra. One more additional
spectrum was obtained with the Subaru telescope using the HDS spectrograph. However, despite its
superb resolution (R $\sim$ 160,000) we cannot separate all four components because it was taken at
a close phase of both binaries, with so much blending that only the more prominent lines of the B
pair were identifiable.

\section{Analysis} 

We conducted a detailed iterative analysis of the system using only TESS and ground-based $VRI$
photometry via the {\sc PHOEBE} analysis package, ver. 0.32svn \citep{2005ApJ...628..426P}. The
other photometric data were used as a complementary material for the derivation of the
minima times for ETV analysis alone. First, the ephemerides were fixed at the values derived from the
preliminary analysis mentioned in Section \ref{sec:data}, effectively removing the ETVs so that
each pair of eclipses could be fit with a purely periodic ephemeris. We then iteratively fit each
of the inner eclipsing photometric signals after removing the eclipses of the other eclipsing pair.
At different time epochs the ephemeris of the pair under analysis is adjusted according to the
phase of the A-B orbit based on the initial AFP analysis. This iterative method was substantiated
by a fact that the two binaries are rather separate and any dynamical interaction is very weak and
slow. For a more compact quadruples, we would need a more sophisticated joint photodynamic
approach; see for example the recent study by \cite{2020arXiv200608979R}.

The values of the third light (i.e., the light contribution of the other pair to the solution of
particular pair) are taken as free parameters. We started with an assumption of a 50/50 light
ratio of the two pairs, and then refit the third light contribution at each step. For the {\sc
PHOEBE} fitting of the light curves we used the following assumptions: the synchronous rotation of
both components (i.e., $F_i=1$), the gravity darkening coefficients, albedos kept fixed at their
suggested values ($g_i=0.32$, $A_i=0.5$), and limb-darkening coefficients being
interpolated from van Hamme's tables \citep{1993AJ....106.2096V}. Differential corrections were
used for fitting along with a $\chi^2$ minimization. This method is exactly the same as
used in our previous work on doubly eclipsing binaries published in \cite{2019A&A...630A.128Z}.

Because of the doubly eclipsing quadruple nature of this system, we used several different approaches
to analyze the spectroscopic data. These were the classical cross-correlation functions CCF
\citep{1979AJ.....84.1511T}, 2D cross-correlation \citep{1994ApJ...420..806Z}, manual cross-correlation
\citep{1996ASPC..101..187S}, and spectral disentangling \citep{2004PAICz..92...15H}. All four
components were not detected in all spectra with each of these methods. Our final RVs were then
computed as an average of these different values as obtained via different techniques. The code
RaveSpan was used for derivation of RVs for some of the spectra \citep{2017ApJ...842..110P}.

\begin{figure*}
  \centering
  \includegraphics[width=0.892\textwidth]{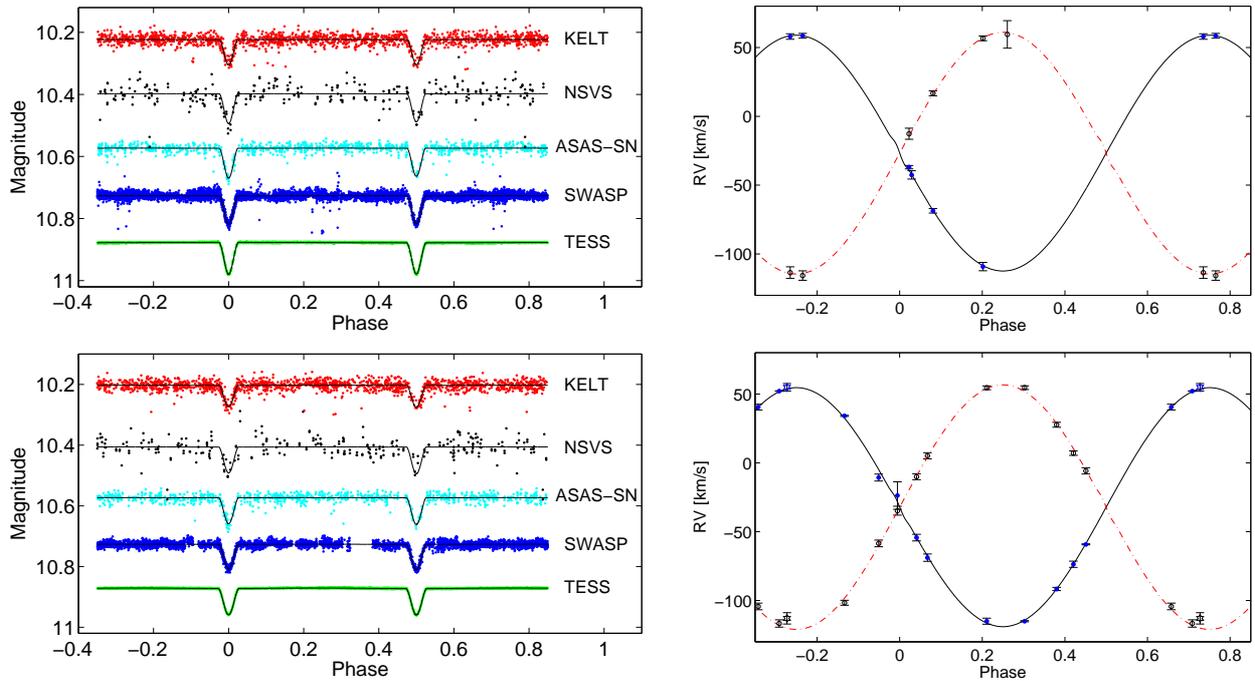}
  \caption{Light curves (left) and RV curves (right) of both pairs A (upper plots) and B (lower). For the RVs, the blue dots and solid line represent the observations and model of the primary pair, while the open circles and the dash-dotted curve represent the secondary components. One spectrum from Subaru was plotted using hexagram symbol.}
  \label{FigLCaRV}
\end{figure*}

\section{Results}  \label{sec:results}

\begin{figure}
  \centering
  \includegraphics[width=0.47\textwidth]{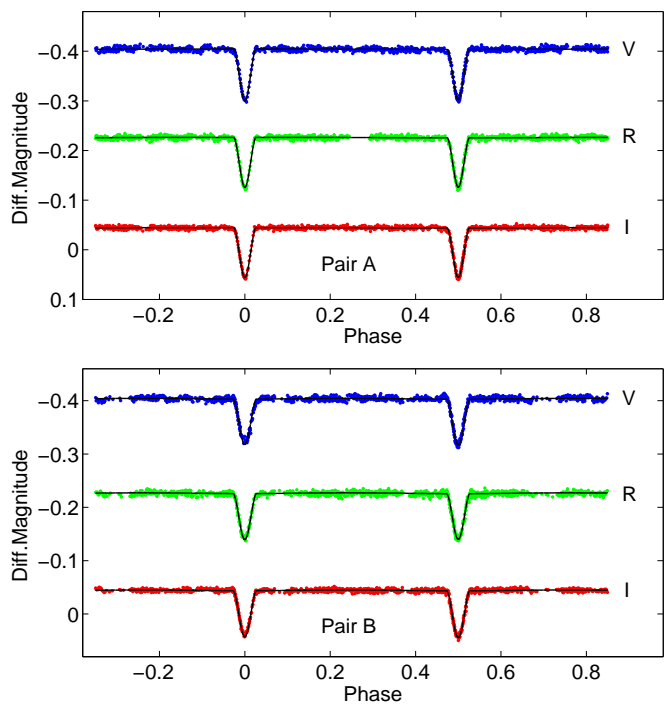}
  \caption{Light curves obtained by Z.H. (Velt\v{e}\v{z}e).}
  \label{FigHenzlLC}
\end{figure}

Complete fits of our light curve and RV data are plotted in Figure \ref{FigLCaRV}, while the  light
curve by Z.H. is plotted in Figure \ref{FigHenzlLC}. The luminosity ratio of pair A to pair B is about
45\%/55\%, as shown in Table \ref{LCRVparam}. Both inner pairs have circular orbits and are
well-detached eclipsing systems. Both eclipsing pairs can be denoted as twins because both inner mass
ratios are very close to unity. Pair B (i.e., the pair with a slightly longer orbital period) seems to
dominate in luminosities and in masses, while its primary component is the brightest and most massive
member of the system. All spectral types are late-F to early-G dwarf stars.

\begin{table}
  \caption{Parameters from the combined fitting of the photometry and RV of both eclipsing pairs, as well as of their mutual orbit.}
  \label{LCRVparam}
  \centering
  \scalebox{0.85}{
  \begin{tabular}{c c c c c}
\hline \hline
                    &\multicolumn{2}{c}{Pair A}                    & \multicolumn{2}{c}{Pair B} \\
Parameter           & Primary   & Secondary                        & Primary   & Secondary    \\
 \hline
 $HJD_0$            & \multicolumn{2}{c}{2455002.4990 $\pm$ 0.0051}& \multicolumn{2}{c}{2455002.5252 $\pm$ 0.0050} \\
 $P$ [d]            & \multicolumn{2}{c}{4.1084261 $\pm$ 0.0000078}& \multicolumn{2}{c}{4.6755210 $\pm$ 0.0000092} \\
 $a$ [R$_\odot$]    & \multicolumn{2}{c}{14.14 $\pm$ 0.12}         & \multicolumn{2}{c}{16.36 $\pm$ 0.10}  \\
 $v_\gamma$ [km/s]  & \multicolumn{2}{c}{-26.73 $\pm$ 1.83}        & \multicolumn{2}{c}{-32.14 $\pm$ 1.58} \\
 $q = M_2/M_1$      & \multicolumn{2}{c}{0.98 $\pm$ 0.02}          & \multicolumn{2}{c}{0.98 $\pm$ 0.02} \\
 $i$ [deg]          & \multicolumn{2}{c}{84.72 $\pm$ 0.09}         & \multicolumn{2}{c}{82.86 $\pm$ 0.08} \\
 $K$ [km/s]         &  85.6 $\pm$ 0.9   & 87.7 $\pm$ 0.8           &  86.9 $\pm$ 0.8   & 88.8 $\pm$ 0.8   \\
 $T$ [K]            &  6000 (fixed)     & 5985 $\pm$ 19            &  6500 (fixed)     & 6494 $\pm$ 15   \\
 $M$ [M$_\odot$]    &  1.14 $\pm$ 0.02  & 1.11 $\pm$ 0.02          &  1.36 $\pm$ 0.02  & 1.33 $\pm$ 0.02 \\
 $R$ [R$_\odot$]    &  1.31 $\pm$ 0.03  & 1.27 $\pm$ 0.02          &  1.71 $\pm$ 0.02  & 1.69 $\pm$ 0.02 \\
 $M_{bol}$ [mag]    &  3.99 $\pm$ 0.03  & 4.06 $\pm$ 0.03          &  3.07 $\pm$ 0.03  & 3.09 $\pm$ 0.03 \\
 $L_{TESS} [\%]$    &  23.3 $\pm$ 0.1   & 21.6 $\pm$ 0.1           &  27.8 $\pm$ 0.1   & 27.3 $\pm$ 0.1  \\
 $L_V [\%]$         &  23.0 $\pm$ 0.8   & 21.4 $\pm$ 1.0           &  28.0 $\pm$ 0.7   & 27.6 $\pm$ 0.9  \\
 $L_R [\%]$         &  23.2 $\pm$ 0.3   & 21.6 $\pm$ 0.4           &  27.6 $\pm$ 0.2   & 27.5 $\pm$ 0.7  \\
 $L_I [\%]$         &  23.3 $\pm$ 0.2   & 21.7 $\pm$ 0.2           &  27.7 $\pm$ 0.2   & 27.3 $\pm$ 0.3  \\
 $P_{AB}$ [yr]      & \multicolumn{4}{c}{34.2 $\pm$ 3.2}\\
 $T_0$ [HJD]        & \multicolumn{4}{c}{2457994 $\pm$ 1109}\\
 $A_{A}$ [d]        & \multicolumn{4}{c}{0.044 $\pm$ 0.003}\\
 $A_{B}$ [d]        & \multicolumn{4}{c}{0.037 $\pm$ 0.003}\\
 $\omega_{AB}$ [deg]& \multicolumn{4}{c}{58.8 $\pm$ 9.0}\\
 $e_{AB}$           & \multicolumn{4}{c}{0.379 $\pm$ 0.017}\\
 \hline
\end{tabular}}
\end{table}

Using these results, we tried to model the mutual orbit of the eclipsing pairs around a common
barycenter to obtain a self-consistent dynamical analysis of the whole system. All available
photometric observations collected (the survey data and our new dedicated observations) were used to
construct the $O-C$ diagrams to detect the ETVs. However, we found our solution from the preliminary
analysis to be problematic. The amplitudes of both ETVs should be in accordance with the derived masses
of both pairs, that is, $A_A/A_B = m_B/m_A$. Even though the O-C analysis finds that the ratio should
be about 2.1, the PHOEBE analysis finds the pair masses to be much more similar to each other, yielding
$A_A/A_B = m_B/m_A \approx 1.2$. Hence, we have to modify our model of the mutual orbit and fix this
ratio of both ETV amplitudes. Of course whole our ETV analysis was redone with the updated ephemerides
and parameters from this latter solution. This final solution is presented in the lower part of Table
\ref{LCRVparam} and is plotted in Figure \ref{FigOC}. As shown, the mutual period is now longer and
still only part of it is covered with data. Both of the ETV amplitudes are also larger. But the very
first data points of pair A now deviate from the fit more than with the previous solution. This is
probably caused by the very poor quality of the NSVS data, both in phase coverage of both minima and
its photometric accuracy. Nevertheless, the mass ratio of both pairs is more realistic and we also
believe that the whole figure is more self-consistent now. The eccentricity of the orbit is not as
high, hence also the periastron passage, which already happened in 2018, did not give us a clear
indication of the whole curvature of ETVs. Additional photometric data is expected to result in a more
accurate determination of the orbital parameters.

\section{Discussion and conclusions}  \label{discussion}

The discovery and analysis of the system CzeV1731 represents a rare configuration of a 2+2
quadruple star system containing two eclipsing binaries, the third system of its kind known to
date. The orbital periods of the inner eclipsing pairs are similar, making this system
different from the two previously known such systems, V994~Her and V482~Per.

The mutual orbital period of the two pairs of about 33~yr is much longer than both inner binary
orbits, hence any mutual dynamical effects \citep{2013ApJ...768...33R} should be very small.
Long-term perturbations such as nodal precession of the orbits should be very slow; the typical order
of magnitude estimation of such an effect $P_{AB}^2 /P_B$ is long, at $\approx 90000~$yr.

\begin{figure}
  \centering
  \includegraphics[width=0.48\textwidth]{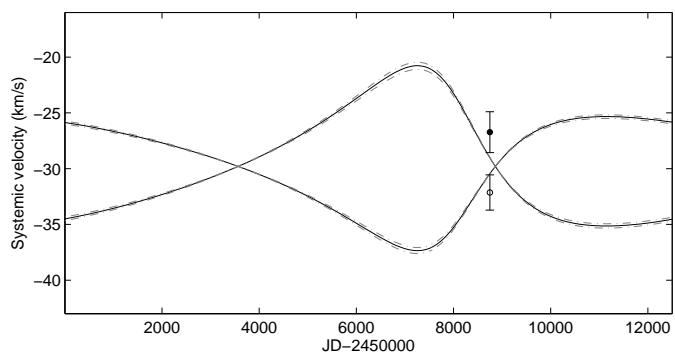}
  \caption{Gamma velocities of both pairs from the RV analysis and the prediction of the systemic velocities on the mutual
  orbit according to our final solution.}
  \label{FigRVs}
\end{figure}

However, a more interesting possibility for future observations is the calculation of the angular
separation of the two pairs on the sky. Owing to its relatively bright nature ($V\sim10.5$), the system
may be suitable for interferometry or even direct imaging. Using the \emph{Gaia}
\citep{2018A&A...616A...1G,2016A&A...595A...1G} distance of 589~pc, the expected angular separation is
about 59~mas. With a small bolometric magnitude difference between the two sources ($<1~$mag) we can
hope CzeV1731 would soon be resolved. Our solution for the outer orbit is still only partly constrained
by data, so further observations would help verify the duration of the outer orbit. The possibility
that the outer orbit is longer than that derived by this work is weakly supported by Figure
\ref{FigRVs}, in which we plotted the two gamma velocities of both inner pairs and their prediction on
the wide orbit. From the parameters of the orbits and individual masses we can calculate the value of
inclination of the orbit as 53.7 $\pm$~2.5~deg. This value can be used to calculate the predicted
systemic velocities and these are plotted in Fig. \ref{FigRVs}. It seems likely that the orbit is
longer and the periastron passage occurs later, but only further observations in the upcoming years
would be able to confirm or rule out this hypothesis.

The GAIA distance itself can also be compared with our estimate of the photometric distance to the
system using our inferred values of parameters. The results of our calculations indicated that the
system is even slightly closer, about 514 $\pm$~11~pc from the Sun. Such a value would even be
better for a prospective interferometric detection because the predicted angular separation
would also be a bit larger (67~mas).

\begin{acknowledgements}
We would like to thank an anonymous referee for his/her useful suggestions improving the level of the
manuscript. The research was supported by the project Progress Q47 Physics of the Charles University in
Prague. This work is supported by MEYS (Czech Republic) under the projects MEYS LM2015046, LTT17006 and
EU/MEYS CZ.02.1.01/0.0/0.0/16$\_$013/0001403. H.Lehmann acknowledges support by DFG grant LE 1102/3-1.
Daniel J. Stevens is supported as an Eberly Research Fellow by the Eberly College of Science at the
Pennsylvania State University. The Center for Exoplanets and Habitable Worlds is supported by the
Pennsylvania State University, the Eberly College of Science, and the Pennsylvania Space Grant
Consortium. We also do thank the SWASP, NSVS, and ASAS-SN teams for making all of the observations
easily public available. Part of the data were collected during the photometric monitoring observations
with the robotic and remotely controlled observatory at the University of Athens Observatory - UOAO.
The observations by Z.H. in Velt\v{e}\v{z}e were obtained with a CCD camera kindly borrowed by the
Variable Star and Exoplanet Section of the Czech Astronomical Society. Simon Albrecht, Teruyuki Hirano,
and Maria Hjorth are greatly acknowledged for obtaining one SUBARU spectrum. This work has made use of
data from the European Space Agency (ESA) mission {\it Gaia} (\url{https://www.cosmos.esa.int/gaia}),
processed by the {\it Gaia} Data Processing and Analysis Consortium (DPAC,
\url{https://www.cosmos.esa.int/web/gaia/dpac/consortium}). Funding for the DPAC has been provided by
national institutions, in particular the institutions participating in the {\it Gaia} Multilateral
Agreement. This research has made use of the SIMBAD and VIZIER databases, operated at CDS, Strasbourg,
France and of NASA Astrophysics Data System Bibliographic Services. This paper includes data collected
with the TESS mission, obtained from the MAST data archive at the Space Telescope Science Institute
(STScI). Funding for the TESS mission is provided by the NASA Explorer Program. STScI is operated by
the Association of Universities for Research in Astronomy, Inc., under NASA contract NAS 5-26555.

\end{acknowledgements}

\end{document}